\documentstyle[12pt,epsf]{article}
\renewcommand{\bar}[1]{\overline{#1}}

\begin{document}

\begin{flushright}
USM-TH-99\\
hep-ph/0009297
\end{flushright}
\bigskip\bigskip

\centerline{\large \bf Nucleon Transversity Distribution }

\centerline{\large \bf from Azimuthal Spin Asymmetry in Pion
Electroproduction}

\vspace{18pt} \centerline{\bf Bo-Qiang Ma
$^{a}$, 
Ivan Schmidt
$^{b}$, 
and Jian-Jun Yang
$^{b,c}$}

\vspace{8pt}

{\centerline {$^{a}$Department of Physics, Peking University,
Beijing 100871, China,}} 

{\centerline {and CCAST (World Laboratory),
P.O.~Box 8730, Beijing 100080, China}}


{\centerline {$^{b}$Departamento de F\'\i sica, Universidad
T\'ecnica Federico Santa Mar\'\i a,}}

{\centerline {Casilla 110-V, 
Valpara\'\i so, Chile}}

{\centerline {$^{c}$Department of Physics, Nanjing Normal
University,}}

{\centerline {Nanjing 210097, China}}

\vspace{6pt}
\begin{center} {\large \bf Abstract}

\end{center}
The azimuthal asymmetry observed by the HERMES
collaboration in semi-inclusive pion production in deep inelastic
scattering of unpolarized positron on the longitudinally polarized
proton target, can provide information of the quark transversity
distributions of the nucleon. We show that the quark transversity
distributions predicted both by the light-cone
quark-spectator-diquark model and by a pQCD inspired model can
give consistent descriptions of the available HERMES data for the
analyzing powers $A_{UL}^{\sin \phi}$ and $A_{UL}^{\sin 2 \phi}$
for $\pi^+$ and $\pi^-$ productions. We also show that the two
models give similar predictions of $A_{UL}^{\sin  \phi}$ for
$\pi^+$ production, whereas they give very different predictions
of  $A_{UL}^{\sin \phi}$ for $\pi^-$ production at large $x$.
Further precision measurement of $A_{UL}^{\sin \phi}$ for
$\pi^-$ production can provide a decisive test of different
models.

\vfill \centerline{PACS numbers: 13.87.Fh, 13.60.-r, 13.88.+e,
14.20.Dh}

\vfill
\centerline{To appear in  Phys. Rev. D}
\vfill

\newpage

Recently, the HERMES collaboration reported evidence for
single-spin asymmetries for semi-inclusive pion production in deep
inelastic scattering (DIS) of unpolarized positron beam on the
longitudinally polarized proton target \cite{HERMES00}. A
significant spin asymmetry of the distribution in azimuthal angle
$\phi$ of the pion relative to the lepton scattering plane is
found. The lepton scattering plane is determined by the incident
and scattered leptons, and the pion emitting plane is determined
by the final detected pion and the virtual photon, with the
virtual photon as the common axis of the two planes. The azimuthal
angle $\phi$ is the angle between the two planes with the axis
direction opposite to the virtual photon. Such azimuthal asymmetry
offers a means to measure the nucleon transversity distributions,
which are one of the three fundamental quark
distributions of the nucleon. The other two
are the unpolarized and helicity distributions, which
are known with some precision both experimentally and
theoretically. The quark transversity distribution is difficult
to be measured, since it is not directly observable in inclusive 
DIS processes. Among the proposals to measure the quark 
transversity distributions, the azimuthal asymmetry in 
semi-inclusive hadron production has been considered 
\cite{Col93,Kot95,Mul96,Jaf98,Ans95},   
through the Collins effect \cite{Col93} of 
non-zero production between a chiral-odd structure function and 
a T-odd fragmentation function. Indeed, there have been a number 
of studies \cite{Kot99,Efr00,Sch00,Bor99,Bog00,Ans00,Suz00} to 
show that the azimuthal asymmetries measured by HERMES can 
provide information concerning the quark transversity 
distributions of the nucleon.

The analyzing power masured by HERMES is defined as
\begin{equation}
A_{UL}^W=\frac{\int \left[{\mathrm d} \phi\right] W(\phi)
\left\{N^+(\phi)-N^-(\phi)\right\}}{\frac{1}{2}\int \left[{\mathrm
d} \phi\right]\left\{ N^+(\phi)+N^-(\phi) \right\}}, \label{AP}
\end{equation}
where $UL$ denotes Unpolarized beam on a
Longitudinally polarized target, $W(\phi)=\sin \phi$ or $\sin 2
\phi$ is the weighting function for picking up the Collins effect,
and $N^+(\phi)$ ($N^-(\phi)$) is the number of events for pion
production as a function of $\phi$ when the target is positively
(negatively) polarized. The analyzing powers for both $\pi^+$ and
$\pi^-$ are measured, and from the data there is clear evidence
for the non-zero values of $A_{UL}^{\sin \phi}$ for $\pi^+$, which
indicate the azimuthal asymmetry. It has been found by Efremov
{\it et al.} \cite{Efr00}, based on a theoretical analysis 
presented in Refs.~\cite{Mul96,Ams}, that the analyzing powers 
$A_{UL}^{\sin \phi}$ and $A_{UL}^{\sin 2 \phi}$ are, 
under a number of simplifying assumptions, proportional 
to the ratios \begin{equation}
\frac{\sum_a e^2_a \delta q^a (x) \left< \delta D^{a/\pi}(z)/z
\right>}{\sum_a e^2_a q^a(x) \left< D^{a/\pi}(z)\right>},
\label{AP1}
\end{equation}
and
\begin{equation}
\frac{ x ^2 \sum_a e^2_a \left( \int_x^1 {\mathrm d} \xi \delta
q^a (\xi) /\xi^2\right) \left< \delta D^{a/\pi}(z)\right>}{\sum_a
e^2_a q^a(x) \left< D^{a/\pi}(z)\right> }, \label{AP2}
\end{equation}
respectively. Here $e_a$ is the charge of the quark with 
flavor $a$, $q^a(x)$ and $\delta q^a(x) $ are the quark 
unpolarized and transversity distributions of the nucleon target,
and $D^{a/\pi}(z)$ and $\delta D^{a/\pi}(z)$ are the 
fragmentation functions to the pion $\pi$ from an unpolarized and
transversely polarized quark with flavor $a$. 
In fact the ratios taken from Ref.~\cite{Mul96} 
involve higher twist functions and
functions with intrinsic transverse momentum, 
and the latter functions are
also related to twist three functions. 
In a next step all these functions
are rewritten into 
twist two distribution and fragmentation functions using
Wandzura-Wilczek type of relations, which are
approximations.
A further detailed 
theoretical analysis can be found in Ref.~\cite{Ans00}. Therefore
with the inputs of $D^{a/\pi}(z)$ and $\delta D^{a/\pi}(z)$ under some
simplifying assumptions, we 
are able to get the quark transversity distributions $\delta 
q^a(x)$ from the measured analyzing powers. We will, following 
Ref.~\cite{Efr00}, consider only the contributions from the 
favored fragmentation functions $D^{a/\pi}$ and $\delta 
D^{a/\pi}$, i.e., $D(z)=D^{u/\pi^{+}}(z)=D^{\bar{d}/\pi^{+}}(z)=
D^{d/\pi^{-}}(z)=D^{\bar{u}/\pi^{-}}(z)$ and similarly for $\delta
D^{a/\pi}$. The average values for $\left< D^{a/\pi}(z) \right>$,
$\left< \delta D^{a/\pi}(z) \right>$, $\left< \delta
D^{a/\pi}(z)/z \right>$, and the corresponding parameters are also
chosen the same as Ref.~\cite{Efr00}. Therefore we only need the 
quark momentum and transversity distributions to calculate
Eqs.~(\ref{AP1}) and (\ref{AP2}).

We now focus our attention on the quark transversity distributions
of the nucleon. It it widely known that the bulk features of the
quark momentum and helicity distributions of the nucleon can be
well described by the quark-spectator-diquark model
\cite{Fey72,DQM,Ma96} and a pQCD based counting rule analysis
\cite{Far75,countingr,Bro95}. Both models have their own
advantages and played important roles in the investigation of
various nucleon structure functions.
However, there are still some unknowns concerning the sea content
of the nucleon and the large $x$ behaviors of valence quarks. For
example, there are still some uncertainties concerning the flavor
decomposition of the quark helicity distributions at large $x$,
especially for the less dominant $d$ valence quark of the proton.
There are two different theoretical predictions of the ratio
$\Delta d(x)/d(x)$ at $x \to 1$: the pQCD based counting rule
analysis \cite{Bro95} predicts $\Delta d(x)/ d(x) \to 1$ whereas
the SU(6) quark-spectator-diquark model \cite{Ma96} predicts
$\Delta d(x)/ d(x) \to -1/3$. There are still no precision
experimental data which can provide a decisive test of the above
two different predictions. In the following analysis we will show
that the same discrepancy also exists concerning the quark
transversity distributions for the $d$ valence quark at large $x$.
We know that the azimuthal spin asymmetry in $\pi^-$ production of
unpolarized lepton DIS scattering on the longitudinally polarized
proton target is sensitive to the quark transversity of the
valence $d$ quark at large $x$, therefore a measurement of
the azimuthal asymmetry of the $\pi^-$ production at large $x$
should be able to provide a decisive test of the different
predictions.

The SU(6) quark-spectator-diquark model \cite{Fey72,DQM,Ma96}
starts from the three quark  SU(6) quark model wavefunction of the
baryon, and if any one of the quarks is probed, re-organize the
other two quarks in terms of two quark wavefunctions with spin 0
or 1 (scalar and vector diquarks), i.e., the diquark serves as an
effective particle which is called the spectator. Some
non-perturbative effects such as gluon exchanges between the two
spectator quarks or other non-perturbative gluon effects in the
hadronic debris can be effectively taken into account by the mass
of the diquark spectator. The mass difference between the scalar
and vector diquarks has been shown to be important for producing
consistency with experimental observations of the ratio
$F_2^n(x)/F_2^p(x)=1/4$ at $x \to 1$ found in the early
experiments \cite{Fey72,DQM}, and also for predicting the 
polarized spin dependent structure functions of the proton and 
the neutron at large $x$ \cite{DQM,Ma96}. The light-cone SU(6)
quark-spectator-diquark model \cite{Ma96} is a revised version of
the same framework, by taking into account the Melosh-Wigner
rotation effect \cite{Ma91b,Sch97},  in order to discuss the quark
helicity and transversity distributions of the nucleon. More
explicitly, the quark helicity and transversity distributions
should be written as \cite{Ma91b,Sch97}
\begin{equation}
\Delta q (x) =\int [{\rm d}^2{\bf k}_{\perp}] M_q(x,{\bf
k}_{\perp}) \Delta q_{QM} (x,{\bf k}_{\perp}); \label{Melosh1}
\end{equation}
\begin{equation}
\delta q (x) =\int [{\rm d}^2{\bf k}_{\perp}] {\hat M}_q(x,{\bf
k}_{\perp}) \Delta q_{QM} (x,{\bf k}_{\perp}), \label{Melosh2}
\end{equation}
where $ \Delta q_{QM} (x,{\bf k}_{\perp})$ is the quark spin
distribution in the quark model, and
$M_q(x,{\bf k}_{\perp})=\frac{(k^+ +m)^2-{\bf k}^2_{\perp}} {(k^+
+m)^2+{\bf k}^2_{\perp}}$ 
and
${\hat M}_q(x,{\bf k}_{\perp})=\frac{(k^+ +m)^2} {(k^+ +m)^2+{\bf
k}^2_{\perp}}$ 
are the corresponding Melosh-Wigner rotation factors due to the
relativistic effect of the quark transversal motion. 
Some further detailed discussions can be found in Refs.~\cite{Ma98a,Ma98b}.

We need to point out that the quark-diquark model with simple
wavefunctions can provide good relations between different
quantities where the uncertainties in the model can be canceled
between each other. It is impractical to expect a good description
of the absolute magnitude and shape for a physical quantity.
However, we may use some useful relations to connect the
unmeasured quantities with the measured quantities. For example,
we may use the following relation to connect the quark
transversity distributions with the quark unpolarized 
distributions 
\begin{equation}
\begin{array}{clcr}
\delta u_{v}(x)
    =[u_v(x)-\frac{1}{2}d_v(x)]\hat{W}_S(x)-\frac{1}{6}d_v(x)\hat{W}_V(x);\\
\delta d_{v}(x)=-\frac{1}{3}d_v(x)\hat{W}_V(x), \label{udv}
\end{array}
\end{equation}
in a similar way as was done for the quark helicity distributions
\cite{Ma96}. We can use the valence quark momentum distributions 
$u_v(x)$ and $d_v(x)$ from one set of quark distribution 
parametrization as inputs to calculate the quark transversity 
distributions, with inputs of $\hat{W}_S(x)$ and $\hat{W}_V(x)$ 
from model calculation \cite{Ma98a}. In this way we can make more
reliable prediction for the absolute magnitude and shape of a 
physical quantity than directly from the model calculation.

We notice that the $d$ quark in the proton is predicted to have a
negative quark helicity distribution at $x \to 1$, and this
feature is different from  the pQCD counting rule prediction of
``helicity retention", which means that the helicity of a valence
quark will match that of the parent hadron at large $x$.
Explicitly, the quark helicity distributions of a hadron $h$ have
been shown to satisfy the counting rule \cite{countingr},
\begin{equation}
q_h(x) \sim (1-x)^p, \label{pl}
\end{equation}
where
$p=2 n-1 +2 \Delta S_z$.
Here $n$ is the minimal number of the spectator quarks, and
$\Delta S_z=|S_z^q-S_z^h|=0$ or $1$ for parallel or anti-parallel
quark and hadron helicities, respectively \cite{Bro95}. Therefore
the anti-parallel helicity quark distributions are suppressed by a
relative factor $(1-x)^2$, and consequently $\Delta q(x)/q(x) \to
1$ as $x \to 1$. Taking only the leading term, we can  write the
quark helicity distributions of the valence quarks as
\begin{equation}
\begin{array}{cllr}
&q^{\uparrow}_{i}(x)=\frac{\tilde{A}_{q_{i}}}{B_3}
x^{-\frac{1}{2}}(1-x)^3;\\
&q^{\downarrow}_{i}(x)=\frac{\tilde{C}_{q_{i}}}{B_5}
x^{-\frac{1}{2}}(1-x)^5,
\end{array}
\label{case2}
\end{equation}
where $\tilde{A}_{q}+\tilde{C}_{q}=N_q$ is the valence quark
number for quark $q$,  $B_n=B(1/2,n+1)$ is the $\beta$-function
defined by $B(1-\alpha,n+1)=\int_0^1 x^{-\alpha}(1-x)^{n} {\mathrm
d} x$ for $\alpha=1/2$, and $B_3=32/35$ and $B_5=512/693$. The
application of the pQCD counting rule analysis to discuss the
unpolarized and polarized structure functions of nucleons can be
found in Ref.~\cite{Bro95}, and the extension to the $\Lambda$ can
be found in Refs.~\cite{MSY23,MSSY57}.

The quark transversity distributions are closely related to the
quark helicity distributions. Soffer's
inequality \cite{Soffer} constrains the quark transversity
distributions by the quark unpolarized and polarized
distributions, and there also exists an approximate relation
\cite{Ma98a}, which connects the quark transversity distributions
to the quark helicity and spin distributions. Two sum rules
\cite{Ma98b}, connecting the integrated quark transversities to
some measured quantities and two model correction factors with
limited uncertainties, have been also recently obtained. For
example, if we assume the saturation of Soffer's inequality
$2 |\delta q(x)| \le  q(x)+ \Delta q(x)$ \cite{Soffer}, 
then we obtain $\delta q=\frac{1}{2}\left[q(x)+\Delta q(x)\right]
=q^{\uparrow}(x)$, and this suggests that in general we may
express $\delta q(x)$ in terms of $q^{\uparrow}(x)$ and
$q^{\downarrow}(x)$. All these considerations indicate that it is
convenient to parameterize the valence quark transversity
distributions in a similar form as the helicity distributions.
Therefore we use as a second model
\begin{equation}
\delta q(x)=\frac{\hat{A}_{q}}{B_3}
x^{-\frac{1}{2}}(1-x)^3-\frac{\hat{C}_{q}}{B_5}
x^{-\frac{1}{2}}(1-x)^5,
\end{equation}
which clearly satisfies Soffer's inequality. 
These
quark transversity distributions are constrained by the 
values of $\delta Q= \int_0^1\delta q(x) {\mathrm d} x $ from the
two sum rules in Ref.~\cite{Ma98b}, and we also use
$\hat{A}_{q}+\hat{C}_{q}=N_q$ as a constraint, just as in the 
case of the helicity distributions, in order to reduce the 
number of uncertain parameters. With the inputs of quark helicity
sum $\Sigma=\Delta U+ \Delta D +\Delta S \approx 0.3$, the 
Bjorken sum rule $\Gamma^p-\Gamma^n=\frac{1}{6}(\Delta U -\Delta 
D)= \frac{1}{6}g_A/g_V \approx 0.2$, both obtained in 
deep-inelastic lepton-nucleon scattering experiments 
\cite{Ma98a,Ma98b}, and taking the two model correction factors 
both to be equal to 1 for the two sum rules of quark 
transversities \cite{Ma98b}, we obtain $\Delta U=0.75$, $\Delta 
D=-0.45$, $\delta U=1.04$, and $\delta D=-0.39$ for the proton, 
assuming $\Delta S=0$. 
We may readjust the values when experimental constraints become 
available, or if we believe other models are more reasonable 
\cite{Ma98b}. The parameters for quark distributions of 
the nucleons and the $\Lambda$ can be found in Table 1.
The ratios $\Delta q(x)/q(x)$ and $\delta q(x)/q(x)$ for the
valence quarks of the proton are predicted to be 1 at $x \to 1$.

\vspace{0.5cm}

\centerline{Table 1~~ The  parameters for quark distributions of
the proton in the pQCD inspired model}

\vspace{0.3cm}

\begin{footnotesize}
\begin{center}
\begin{tabular}{|c||c|c||c|c|c|c||c|c|c|c|}\hline
 Baryon & $q_1$ & $q_2$ & $\tilde{A}_{q_1}$
 &$\tilde{C}_{q_1}$ &$\tilde{A}_{q_2}$
 &$\tilde{C}_{q_2}$ & $\hat{A}_{q_1}$
 &$\hat{C}_{q_1}$ &$\hat{A}_{q_2}$
 &$\hat{C}_{q_2}$ \\ \hline
p & u & d & 1.375 & 0.625 & 0.275 & 0.725 &1.52&0.48&0.305&0.695\\
\hline
\end{tabular}
\end{center}
\end{footnotesize}
\vspace{0.5cm}

\begin{figure}
\begin{center}
\leavevmode {\epsfysize=5cm \epsffile{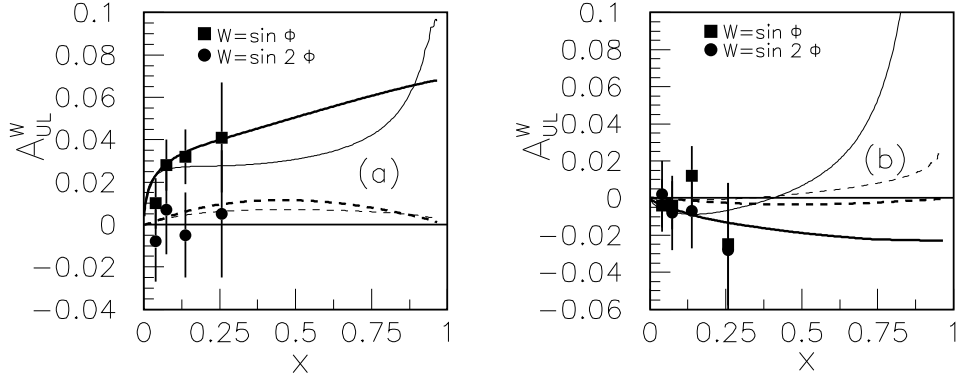}}
\end{center}
\caption[*]{\baselineskip 13pt The analyzing powers $A_{UL}^{\sin
\phi}$ and $A_{UL}^{\sin 2 \phi}$ for (a) $\pi^+$ and (b) $\pi^-$
semi-inclusive production in deep inelastic scattering of
unpolarized positron on the longitudinally polarized proton
target. The thick and thin curves correspond to the calculated
results from the light-cone quark-diquark model
and the pQCD inspired model respectively, with the solid and
dahsed curves corresponding to $A_{UL}^{\sin \phi}$ and
$A_{UL}^{\sin 2 \phi}$ respectively. The inputting quark
distributions are from CTEQ5 set 1 parametrization \cite{CTEQ5} at
$Q^2=4$~GeV$^2$. }\label{msy9f1}
\end{figure}

In the denominators of Eqs.~(\ref{AP1}) and (\ref{AP2}), there are
contributions from both quark and antiquark fragmentations, and
we should take them into account at small $x$. Using the CTEQ
parametrization of quark distributions \cite{CTEQ5}, we
can calculate the contributions from quarks and antiquarks for the
unpolarized quark distributions. We also use the values 
of $u_v(x)$ and $d_v(x)$ from the CTEQ parametrization as the 
inputs in Eq.~(\ref{udv}) to calculate the quark transversity 
distributions of the valence quarks. This is consistent with the 
calculation of the denominators, since we have the same inputs. 
In Fig.~\ref{msy9f1} we present the calculated $A_{UL}^{\sin 
\phi}$ and $A_{UL}^{\sin 2 \phi}$ for both $\pi^+$ and $\pi^-$
productions with only the valence quark distributions in the
numerators of Eqs.~(\ref{AP1}) and (\ref{AP2}). We can see from
Fig.~\ref{msy9f1} that the calculated results are compatible with
the HERMES data of the analyzing powers, and this is in agreement
with existing interpretations of single-spin asymmetries as being
associated with the valence quarks distributions \cite{Art97,ABM}.
In Fig.~\ref{msy9f1} we also present the calculated results with
the quark transversity distributions from the pQCD inspired model,
and find that the results are also in compatible with the existing
data.

We notice that the quark-diquark model and the pQCD inspired model
give similar predictions of the analyzing power $A_{UL}^{\sin
\phi}$ for $\pi^+$ production, whereas they give very
different predictions of $A_{UL}^{\sin \phi}$ for  $\pi^-$
production. For $\pi^+$ production, the contributions of the quark
transversity distributions are dominated by the $u$ quarks from
the fragmentation of $u \to \pi^+$. From Table 1 of
Ref.~\cite{Ma98b}, we find that the quark-diquark model and
the two sum rules have similar predictions of $\delta U$, and also
in both the quark-diquark model and the pQCD inspired model, the
$u$ quarks are totally positively polarized inside the proton at
$x \to 1$. Therefore the analyzing power $A_{UL}^{\sin \phi}$ for
the $\pi^+$ production is not sensitive to different models.
However, this situation is quite different for $\pi^-$ production.
We find from Fig.~\ref{msy9f1}(b) that the quark-diquark
model and the pQCD inspired model have different predictions of
$A_{UL}^{\sin \phi}$ for $\pi^-$ production in the region of
$x \ge 0.3$, right above the available experimental data. This can
be easily understood since the quark-diquark model predicts
$\delta d(x)/d(x) =-1/3$ at $x \to 1$, whereas the pQCD inspired
model predicts $\delta d(x)/d(x) =1$. For $\pi^-$ production, the
contributions of the quark transversity distributions are
dominated by the $d$ quark from the fragmentation of $d \to
\pi^-$. Therefore $A_{UL}^{\sin \phi}$ for $\pi^-$ production
goes in opposite directions at large $x$ in the two models.
This prediction does not suffer from any model uncertainty which
might change the calculated results quantitatively. Thus further
precision measurement of $A_{UL}^{\sin \phi}$ for the $\pi^-$
production by HERMES or other groups can provide a decisive test
of the two different predictions. However, 
there should be some uncertainties
in the quantitative predictions. The uncertainties 
for the quark-diquark model
should be small, whereas the uncertainties for 
the pQCD based model are comparatively large. Further constraints
from more experimental data on the pQCD based model 
with higher order terms  
can improve the situation and increase our predictive power for
the pQCD based analysis.

In conclusion, we showed in this paper that the azimuthal
asymmetry observed by the HERMES collaboration in semi-inclusive
pion production in deep inelastic scattering of unpolarized
positron on the longitudinally polarized proton target, can
provide information of the quark transversity distributions of the
nucleon. We showed that the quark transversity distributions
predicted both by the light-cone quark-spectator-diquark model and
by a pQCD inspired model can give consistent descriptions of the
available HERMES data of the analyzing powers $A_{UL}^{\sin \phi}$
and $A_{UL}^{\sin 2 \phi}$ for both $\pi^+$ and $\pi^-$
production. We also showed that the two models give similar
predictions of $A_{UL}^{\sin  \phi}$ for $\pi^+$ production,
whereas they give very different predictions of  $A_{UL}^{\sin
\phi}$ for $\pi^-$ production at large $x$. Further precision
measurement of $A_{UL}^{\sin \phi}$ for $\pi^-$ production can
provide a decisive test of different models.

{\bf Acknowledgments: } This work is partially supported by
National Natural Science Foundation of China under grants
19975052 and 19875024, by Fondecyt (Chile) postdoctoral fellowship
3990048, and by Fondecyt (Chile) grants
1990806 and 8000017, and by a C\'atedra
Presidencial (Chile).

\end{document}